\documentclass[prd,twocolumn,showpacs,superscriptaddress,nofootinbib]{revtex4}
\usepackage{graphicx}

\newcommand{\tran}{{\scriptscriptstyle\top}}
\newcommand{\si}{\mathcal{S}}
\newcommand{\Zsi}{\mathcal{Z}_{\mathcal{S}}}

\newcommand{\tsfrac}[2]{{\textstyle\frac{#1}{#2}}}

\newcommand{\exch}[1]{\vbox{\mathsurround=0pt 
	\ialign{##\crcr$\hss\rlap{$\scriptscriptstyle\rightharpoonup$}\relax
		\hbox{$\scriptscriptstyle\leftharpoonup$}\hss$\crcr
	\noalign{\nointerlineskip}$\displaystyle{#1}$\crcr}}\relax
	\vphantom{#1}}
\begin{document}

\title{Towards a novel wave-extraction method for numerical relativity.\\  I.  Foundations and initial-value formulation}
\author{Christopher~Beetle}
\affiliation{Department of Physics, Florida Atlantic University, Boca Raton, Florida 33431}
\affiliation{Department of Physics, University of Utah, Salt Lake City, Utah 84112}
\author{Marco~Bruni}
\affiliation{Institute of Cosmology and Gravitation,
	Mercantile House, Hampshire Terrace, PO1 2EG, Portsmouth UK}
\author{Lior~M.~Burko}
\altaffiliation[New address: ]{Department of Physics and Astronomy, 
	Bates College, Lewiston, Maine 04240}
\affiliation{Department of Physics, University of Utah, Salt Lake City, Utah 84112}
\author{Andrea~Nerozzi}
\affiliation{Institute of Cosmology and Gravitation,
	Mercantile House, Hampshire Terrace, PO1 2EG, Portsmouth UK}
\affiliation{Center for Relativity, Department of Physics, University of Texas at Austin, 
	Austin, Texas 78712-1081}

\date{\today}
\pacs{04.25.Dm, 04.30.Nk, 04.70.Bw}

\begin{abstract}
The Teukolsky formalism of black hole perturbation theory describes weak gravitational radiation generated by a mildly dynamical hole near equilibrium.  A particular null tetrad of the background Kerr geometry, due to Kinnersley, plays a singularly important role within this formalism.  In order to apply the rich physical intuition of Teukolsky's approach to the results of fully non-linear numerical simulations, one must approximate this Kinnersley tetrad using raw numerical data, with no \textit{a priori} knowledge of a background.  This paper addresses this issue by identifying the directions of the tetrad fields in a quasi-Kinnersley frame.  This frame provides a unique, analytic extension of Kinnersley's definition for the Kerr geometry to a much broader class of space-times including not only arbitrary perturbations, but also many examples which differ non-perturbatively from Kerr.  This paper establishes concrete limits delineating this class and outlines a scheme to calculate the quasi-Kinnersley frame in numerical codes based on the initial-value formulation of geometrodynamics.
\end{abstract}

\maketitle

\section{Introduction}

After an early history marked by vigorous debate (see, e.g., \cite{Khrr}), the existence of gravitational radiation has become accepted as a hallmark prediction of Einstein's general theory of relativity.  Despite indirect evidence supporting this prediction, however, gravitational waves have yet to be observed directly.  Indeed, rough, order-of-magnitude calculations indicate a typical passing wave will produce only tiny material strains of order $10^{-21}$.  A new generation of laser-interferometric gravitational wave observatories have recently been commissioned, and several are already operational, which aim to observe such strains \cite{RHlr} and thereby detect gravitational waves incident on Earth.  These experiments are necessarily extremely delicate and, as a result, both initial detection and the long-term goal of extracting new information about the distant sources of a particular gravitational wave must be aided by detailed theoretical predictions.

A community of theorists have turned to numerical simulations of the full, non-linear Einstein equations to identify the characteristic features of the gravitational radiation generated by various sources.  An impressive array of techniques have developed within this young field of numerical relativity which aim to provide the ongoing experimental effort with accurate predictions.  However, the program faces a number of challenges, from foundational questions to issues of implementation and interpretation of results.  Here, we consider one issue of interpretation.  We ask how, exactly, the well-known theoretical description of gravitational radiation outside quiescent black holes due to Teukolsky \cite{Teq} might be applied to the results of numerical simulations.

The notion of gravitational radiation in general relativity does not have universal meaning.  Rather, a proper definition can be given only in regions of space-time, \textsl{radiation zones}, whose geometry is characterized by two distinct length scales \cite {Tgr}: one describing an ``average'' radius of curvature and a second, much shorter scale corresponding to the wavelength of gravitational waves.  Because these two scales can be distinguished only when the waves contribute small corrections to the ``average'' curvature, many analyses of gravitational radiation are founded on perturbation theory.  Examples include not only the standard analysis in linearized gravity, but the Regge--Wheeler \cite{RWsch} and Zerilli \cite{Zsch} approaches, as well as Teukolsky's.  Even the asymptotic formulation of the gravitational radiation \cite{Arm}, which applies to quite general space-times, operates by studying differences between the physical metric and a fixed asymptotic metric near conformal infinity.  In numerical relativity, however, all of these analyses are difficult to implement.  The perturbation approaches are complicated because no background metric on space-time is known \textit{a priori}, while the asymptotic approach needs access to field values far outside the computational domain.

This paper focusses on Teukolsky's perturbative approach since it describes radiation fields near a rotating black hole, the expected end-state of many physical processes producing relatively strong gravitational wave signals.  We first identify the essential elements of the Teukolsky formalism which derive from its fixed background geometry.  In particular, we are interested in the preferred tetrad of basis vectors on space-time which underlies the definition of Teukolsky's fields.  Although this tetrad arises as a perturbation of a canonical, Kinnersley tetrad \cite{Ktd} for the background Kerr geometry, we show one can approximate it using only the \textit{physical} metric, eliminating any reference to the background.  The construction of this approximate Kinnersley tetrad occurs naturally in two stages.  The first, which is completed in this paper, fixes the directions of the desired tetrad vectors.  Because the final results of our analysis make no mention of a background geometry, they may be applied unambiguously to a broad class of space-times.  We give concrete criteria characterizing this class, and find it includes many examples which differ \textit{non-perturbatively} from the Kerr geometry.  In particular, when a numerical simulation produces a single black hole settling down to equilibrium, this first stage should be viable even at relatively early times after the hole forms.  The problem of the second stage, fixing the scalings of the tetrad vectors along these preferred directions, is described in some detail here but not solved.  We plan to present a solution in a future paper.  However, even the first stage alone provides significant gauge-invariant information partially characterizing the gravitational field encoded in the numerical variables.  An earlier Letter \cite{BBscal} presented one scheme to extract such information, and this paper can be regarded as an elaboration and expansion of that presentation.

Like the Teukolsky formalism itself, several of this paper's results are most simply expressed in the Newman--Penrose approach \cite{NP} to space-time geometry.  However, because many existing numerical codes do not currently implement this approach, we strive to present our final results in a form which does not refer explicitly to Newman--Penrose quantities.  Rather, we seek a form which would allow the Teukolsky fields to be deduced \textit{ab initio} from the physical metric.  There is a price to be paid for this, as certain results appear more complicated in this language.  A complementary, and slightly more comprehensive, analysis based on the Newman--Penrose approach is given in a companion paper \cite{comp}.  This complementary analysis operates by picking an arbitrary tetrad on a generic space-time and showing how to transform it to yield one of interest in the Teukolsky formalism.  We present the two analyses separately to maximize the clarity of each.

The following is an outline of this paper.  First, we review elements of the Teukolsky formalism from the traditional point of view, where a preferred background Kerr geometry is known \textit{a priori}, and formulate in detail the problem to be solved.  Second, we introduce the notion of a \textsl{transverse frame}.  Several such frames exist in a general space-time, all are calculable from the physical geometry, and one is the \textsl{quasi-Kinnersley frame} of interest in the Teukolsky formalism.  Finally, to help deploy these results for numerical relativity, we express these results using only initial-value data on an arbitrary spatial slice.  We conclude with several comments.  Finally, an Appendix elaborates one of the central calculations of the paper.

\section{The Teukolsky Formalism and Numerical Relativity}

The observable signature of a non-trivial gravitational field in general relativity is space-time curvature.  However, because no coordinate system is preferred over any other, the coordinate components of the curvature tensor do not generally have any invariant physical meaning.  One must be careful to distinguish, for example, between actual gravitational radiation and oscillations in those components caused by a peculiar choice of coordinates.  The Teukolsky formalism of first-order perturbation theory \cite{Teq} solves this problem using a family of gauge-invariant\relax  
\footnote{Throughout this section, ``gauge'' refers to the first-order diffeomorphisms of linearized gravity.}
scalar variables originally proposed by Newman and Penrose \cite{NP}.

The Newman--Penrose approach introduces an initially arbitrary \textsl{null tetrad} $(\ell^a, n^a, m^a, \bar m^a)$ on space-time.  The four vector fields here are all null, with the first pair real and the second both complex and conjugate to one another.  Their only non-vanishing inner products are 
\begin{equation}\label{norm}
	\ell^a\, n_a = -1 \qquad\mbox{and}\qquad m^a\, \bar m_a = 1.
\end{equation}
It is convenient to restrict attention to \textsl{oriented} null tetrads satisfying 
\begin{equation}\label{orient}
	24i\, \ell_{[a}\, n_b\, m_c\, \bar m_{d]} = \epsilon_{abcd}, 
\end{equation}
where $\epsilon_{abcd}$ denotes the usual space-time volume element.  This restriction can always be made without loss of generality in cases of physical interest.  The independent components of the Weyl curvature tensor in such a non-coordinate basis are encoded in five complex \textsl{Weyl scalars} 
\begin{equation}\label{Psi}
  \begin{array}{@{}r@{}l@{}}
    \Psi_0 &{}:= C_{abcd} \, \ell^a   \, m^b \, \ell^c   \, m^d \\[.5ex]
    \Psi_2 &{}:= C_{abcd} \, \ell^a   \, m^b \, \bar m^c \, n^d \\[.5ex]
    \Psi_4 &{}:= C_{abcd} \, \bar m^a \, n^b \, \bar m^c \, n^d. 
  \end{array}\quad
  \begin{array}{@{}r@{}l@{}}
    \Psi_1 &{}:= C_{abcd} \, \ell^a   \, m^b \, \ell^c   \, n^d \\[.5ex]
    \Psi_3 &{}:= C_{abcd} \, \ell^a   \, n^b \, \bar m^c \, n^d 
  \end{array}
\end{equation}
These scalar quantities are naturally independent of any space-time coordinate system, but do clearly depend on the choice of null tetrad.  Accordingly, the $\Psi_n$ can acquire direct (observer-independent) physical meaning only when a physically preferred null tetrad can be found.

The Teukolsky formalism picks a particular null tetrad for the Newman--Penrose scalars by exploiting special features of its given Kerr background.  Specifically, any Kerr geometry admits exactly two repeated principal null directions\relax
\footnote{The braces $\{ \cdot \}$ emphasize that only the directions of these vectors are determined.}
$\{ \ell^a \}$ and $\{ n^a \}$.  In any null tetrad $(\ell^a, n^a, m^a, \bar m^a)$ whose real null vectors lie in these directions, all the Newman--Penrose curvature components except  $\Psi_2$ vanish, guaranteeing, as we discuss further below, the gauge-invariance of first-order perturbations in $\Psi_0$ and $\Psi_4$.  It is therefore quite natural to choose the directions of the tetrad fields in this way, but one must proceed to fix their scalings along these directions as well.  A technique to do this in an exact Kerr, or generally in any Petrov type D, geometry was first proposed by Kinnersley \cite{Ktd}.  To explain this technique, it is useful to recall the generic ambiguity in the null tetrad.

At a given point of a general space-time, one can transform any given oriented null tetrad $(\ell^a, n^a, m^a, \bar m^a)$ to any other using elementary transformations of three basic types.  These include two families of elementary \textsl{null rotations} $L(a)$ and $N(b)$, with $a$ and $b$ arbitrary complex parameters, which preserve either $\ell^a$ or $n^a$, respectively, while changing the directions of all three of the other tetrad elements.  A third family of \textsl{spin-boost} transformations $S(c)$, with $c$ an arbitrary complex parameter, mutually scales the real null vectors $\ell^a$ and $n^a$, leaving their directions unchanged, and rotates the complex elements $m^a$ and $\bar m^a$ by complementary phases.  In addition to these, we highlight a single fourth transformation, the \textsl{exchange} operation $E$, which interchanges the real and complex tetrad elements in pairs.  The detailed formulae describing these transformations are unimportant here.  They are standard and can be found, for instance, in \cite{exact, comp}.  We do, however, note the effects of these transformations on the Weyl scalars $\Psi_n$:
\begin{equation}\label{trans}
	\begin{array}{@{}r@{}l@{}}
		L(a) : \Psi_n \mapsto \hat\Psi_n &{}= 
			\sum_{m=0}^n\, {n \choose m}\, a^{n-m}\, \Psi_m \\[1ex]
		N(b) : \Psi_n \mapsto \check\Psi_n &{}= 
			\sum_{m=n}^4\, {{4-n} \choose {m-n}}\, b^{m-n}\, \Psi_m \\[1ex]
		S(c) : \Psi_n \mapsto \tilde\Psi_n &{}= c^{2 - n} \, \Psi_n \\[1ex]
		E: \Psi_n \mapsto \exch\Psi_n &{}= \Psi_{4-n}
	\end{array}
\end{equation}
where $n \choose m$ denote the usual binomial coefficients.

In the Kerr geometry, fixing $\ell^a$ and $n^a$ to lie in the two repeated principal null directions eliminates the null rotations completely, but does not restrict the spin-boost and exchange freedom in any way.  To proceed, Kinnersley implicitly breaks the exchange symmetry by choosing $\ell^a$ to point outward, toward infinity, in the exterior of the hole and then sets 
\begin{equation}\label{NPeps}
	\epsilon := \tsfrac{1}{2}\, \ell^a\, (\ell^b\, \nabla_a\, n_b - m^b\, \nabla_a\, \bar m_b) = 0.
\end{equation}
The quantity $\epsilon$ is one of the standard Ricci rotation coefficients used in the Newman--Penrose approach.  Setting it to zero does not eliminate the residual spin-boost ambiguity entirely, however, as one remains free to do spin-boost transformations with parameters $c$ satisfying
\begin{equation}\label{cKinn}
	\ell^a\, \nabla_a\, c = 0.
\end{equation}
Note that, since the right side vanishes, this equation is independent of the scaling of the vector field $\ell^a$, and implies only that $c$ must be constant along each of its integral curves.  The set of such curves rule the exterior region of the black hole and each intersects a unique point of future null infinity.  As a result, adapting the scaling of the tetrad near infinity to stationary observers there uniquely determines the Kinnersley tetrad everywhere on Kerr space-time.  That is, fixing the scalings of $\ell^a$ and $n^a$ such that their spatial projections relative to a stationary observer have equal magnitude near infinity breaks the remaining spin-boost invariance.

In the Teukolsky formalism, one considers only space-times describing perturbations of a \textit{known} Kerr metric.  The physical metric, consisting of both background and perturbation, generally does not have repeated principal null directions, and does not admit a preferred Kinnserley tetrad.  However, it is nonetheless natural to restrict attention to null tetrads of this metric which differ from the  background's Kinnersley tetrad only at first order in perturbation theory.  This is Teukolsky's approach.  In such a \textsl{perturbed Kinnersley tetrad}, all of the curvature components except $\Psi_2$ vanish at leading order, though all may acquire first-order corrections.  As a result, all components except $\Psi_2$ are invariant under first-order diffeomorphisms, the gauge transformations of the linearized theory.  Moreover, the residual, first-order ambiguity in the tetrad is generated by null rotations with parameters $a$ and $b$ which vanish at leading order, and spin-boosts with parameters $c$ equal to unity at leading order.  The exchange operation, being finite, is disallowed.  These residual transformations certainly modify the perturbed tetrad, but do not affect the values of the curvature components $\Psi_0$, $\Psi_2$ and $\Psi_4$.  Thus, $\Psi_0$ and $\Psi_4$ evaluated in \textit{any} such perturbed tetrad are invariant in both senses, and acquire physical meanings.  Specifically, using an analysis of the geodesic deviation equation due to Szekeres \cite{Sgc, comp}, they are associated with transverse gravitational radiation propagating along the null directions $n^a$ and $\ell^a$, respectively.

The components $\Psi_1$ and $\Psi_3$ in a perturbed Kinnersley tetrad are invariant under infinitesimal diffeomorphisms, but may be given any value whatsoever by performing suitable first-order null rotations on the tetrad.  They are therefore pure gauge in the Teukolsky formalism.  In a variety of situations, including both the reconstruction of metric perturbations in the first-order theory \cite{Chandra} and a second-order perturbation analysis \cite{CLso}, it is convenient to use the remaining null rotation freedom to set 
\begin{equation}\label{Kinn}
	\Psi_1 = 0 = \Psi_3.
\end{equation}
This can always be done, and exhausts the first-order null rotation freedom in the tetrad.  There is no similar convenient criterion to eliminate the first-order spin-boosts, but none is needed in  the Teukolsky formalism. 

Let us now focus on the subtleties inherent in applying Teukolsky's techniques in numerical relativity.  Black holes in general relativity are known to be stable \cite{RWsch, PTstab}.  Thus, when a numerical evolution produces a space-time containing a single black hole, one should expect it to settle down over time, approaching a quiescent final state.  A description via perturbation theory should accordingly become increasingly viable over time, although in practice it may be far from clear exactly \textit{how} a given physical metric should be so described.  One natural response to this difficulty would be to search among \textit{all} possible Kerr metrics on the given space-time for one in which the perturbation, the difference between the physical metric and that particular Kerr metric, is smallest.  However, even if it were clear how to conduct such a search systematically, it would likely be an extremely difficult procedure.  We take a different approach here.  The key element of the Teukolsky formalism provided by the background Kerr metric is its Kinnersley tetrad.  Rather than search for an approximate Kerr metric on space-time whose Kinnersley tetrad will allow us to exploit the Teukolsky formalism, we search directly for an approximation to that Kinnersley tetrad.

Given \textit{only} the physical metric on a perturbed Kerr space-time, a natural way to begin searching for its perturbed Kinnersley tetrad is to seek tetrads for the physical metric satisfying Eq.~(\ref{Kinn}).  That is, to find a tetrad with $\Psi_1$ and $\Psi_3$ vanishing at leading order, simply demand they vanish to all orders.  This eliminates only gauge degrees of freedom in first-order perturbation theory, and there is no obvious problem of principle seeking such tetrads at an arbitrary point of a general space-time.  The components $\Psi_0$ and $\Psi_4$ in such a frame cannot generally be set to zero simultaneously by further refining the choice of tetrad.  Even in perturbation theory, these components are physical and provide a rough, quantitative measure of how ``far'' such a space-time is from Kerr \cite{BBscal, BC}.  Motivated by their association with transverse radiation in perturbation theory, we refer to tetrads satisfying Eq.~(\ref{Kinn}) as \textsl{transverse null tetrads}.  We emphasize, however, that in regions of space-time supporting strongly self-gravitating fields, there may be no precise relation between these tetrads and any proper notion of gravitational radiation.

Even in the comparatively simple case of a perturbed Kerr space-time, there may be many different transverse tetrads.  These may differ from one another in two ways.  First, the transversality condition of Eq.~(\ref{Kinn}) is invariant under the spin-boost and exchange transformations of Eq.~(\ref{trans}) whence, whenever one transverse tetrad exists, so do infinitely many others.  In addition to this generic multiplicity, however, there could exist distinct transverse tetrads on a given space-time related by non-trivial null rotations.  This second ambiguity raises a more serious problem in our context since even the \textit{directions} of the real null vectors in a transverse tetrad on a perturbed Kerr space-time may have nothing to do with the underlying Kinnersley tetrad.  The current paper addresses this second difficulty.  It constructs, \textit{up to spin-boost and exchange}, a unique tetrad on a broad class of space-times comprising a \textit{non-perturbative} neighborhood of the Kerr geometry in the space of solutions to Einstein's equations.  We refer to the product of this construction as the \textsl{quasi-Kinnersley frame}.  Mathematically, it is an equivalence class $\{ \ell^a, n^a, m^a, \bar m^a \}$ of null tetrads which can be transformed into one another using only spin-boost and exchange transformations.  The set of all tetrads at a point is a disjoint union of such equivalence classes, and we refer to these generally as \textsl{null frames}.  Although this nomenclature may be slightly unfortunate --- the terms ``null tetrad'' and ``null frame'' are often used interchangeably in the literature --- we will adhere strictly to it here since the notion of a frame is essential to our argument.

\section{The Quasi-Kinnersley Frame}

A null frame is completely determined by the common \textit{directions}, $\{ \ell^a \}$ and $\{ n^a \}$, of the real null vectors in its constituent tetrads.  These real null vectors span a time-like 2-plane in the tangent space at each point of space-time, while the complex vectors span the orthogonal space-like 2-plane.  Spin-boost and exchange operations may change the bases defined by the tetrad on these two sub-spaces, but not the sub-spaces themselves.  Since knowledge of either orthogonal sub-space determines the other, the real null directions suffice to determine the frame.  Mathematically, we can exploit this interrelation by associating a 2-form\relax
\footnote{Here, and throughout this paper, our convention is that round (square) brackets on indices denote the (anti-)symmetric \textit{part} of a tensor.  That is, $T_{ab} = T_{(ab)} + T_{[ab]}$.}
\begin{equation}\label{sig}
	\Sigma_{ab} = \ell_{[a}\, n_{b]} - m_{[a}\, \bar m_{b]}.
\end{equation}
to any given null tetrad $(\ell^a, n^a, m^a, \bar m^a)$.  This complex 2-form is invariant under spin-boosts and only changes sign under exchange.  In addition, it is self-dual\relax
\footnote{Although self-dual 2-forms are common in the mathematical relativity literature, they are seldom used in numerical work.  A summary of the relevant mathematics may be found in Chapters 3 and 4 of \cite{exact}.  Note, however, that the notation of this paper differs in a few details.  We have therefore endeavored to make the presentation here as self-contained as possible.}
and unit:
\begin{equation}\label{sd}
	{}^\star \Sigma_{ab} := \tsfrac{1}{2}\, \epsilon_{abcd}\, \Sigma^{cd} = +i\, \Sigma_{ab}
	\quad\mbox{and}\quad
	\Sigma_{ab}\, \Sigma^{ab} = -1.
\end{equation}
In fact, \textit{all} unit self-dual 2-forms arise in this way.  For each, one can find an oriented null tetrad of vectors $(\ell^a, n^a, m^a, \bar m^a)$ such that $\Sigma_{ab}$ takes the form of Eq.~(\ref{sig}).  For a given $\Sigma_{ab}$, we have
\begin{equation}\label{recon}
	\begin{array}{@{}r@{}l@{}r@{}l}
		\Sigma_{ab}\, \ell^b &= -\tsfrac{1}{2}\, \ell_a &\qquad
		\Sigma_{ab}\, n^b &= \tsfrac{1}{2}\, n_a \\[1ex]
		\Sigma_{ab}\, m^b &= -\tsfrac{1}{2}\, m_a &\qquad
		\Sigma_{ab}\, \bar m^b &= \tsfrac{1}{2}\, \bar m_a.
	\end{array}
\end{equation}
That is, $\ell^a$ and $n^a$ are the unique \textit{real} null eigenvectors of $\Sigma_{ab}$, while $m^a$ and $\bar m^a$ are its unique complex null eigenvectors orthogonal to both $\ell^a$ and $n^a$.  Every unit, self-dual 2-form has such eigenvectors, each of which is determined only up to scaling, whence the tetrad is not unique.  Rather, there is a canonical one-to-one correspondence between null frames and non-null directions in the complex, three-dimensional space of self-dual 2-forms at each point of space-time.

A non-null, self-dual 2-form $\Sigma_{ab}$ is associated to a \textit{transverse} null frame if and only if 
\begin{equation}\label{eigen}
	C_{ab}{}^{cd}\, \Sigma_{cd} = -4\Psi_2^\tran\, \Sigma_{ab},
\end{equation}
where $C_{ab}{}^{cd}$ denotes the Weyl curvature tensor of space-time.  For most $\Sigma_{ab}$, the right side would include additional terms of the form $4\Psi_1\, \bar m_{[a}\, n_{b]}$ and $4\Psi_3\, \ell_{[a}\, m_{b]}$.  It is straightforward to show that whenever $\Sigma_{cd}$ is self-dual on the left side of Eq.~(\ref{eigen}), the right side is self-dual as well.  Thus, the search for transverse frames reduces to an eigenvalue equation for the Weyl tensor, viewed as a map from the space of self-dual 2-forms to itself.  Notably, the eigenvalue for a given eigen-form of the Weyl tensor is directly connected to the Newman--Penrose scalars in the associated null frame: $\Psi_2^\tran$ denotes the $\Psi_2$ scalar, which is invariant under both spin-boost and exchange, evaluated in a \textit{transverse} frame.

The above result establishes a one-to-one correspondence between transverse frames $\{ \ell^a, n^a, m^a, \bar m^a \}$ and non-null, self-dual eigen-forms $\Sigma_{ab}$ of the Weyl tensor.  The number and nature of the these eigen-forms are very well understood \cite{exact}, and depend solely on the algebraic (Petrov) class of space-time.  In particular, there are exactly three transverse frames at each point of a algebraically general (type I) space-time.  Meanwhile, the Weyl tensors of type D space-times, such as the Kerr solution, have one degenerate eigen-space, leading to infinitely many transverse frames, as well as a distinguished non-degenerate eigen-form associated with the Kinnersley frame.  Type II geometries admit only one transverse frame.  In each of these three cases, therefore, the possible ambiguity in the transverse frame of interest is \textit{finite}.  One can recognize the Kinnersley frame of a type D or type II geometry immediately, and the only ambiguity occurs in the general case, where there are only three possible choices.  The most specialized algebraic classes, types N and III (as well as type 0, where the Weyl tensor vanishes), are very rare and correspond to highly specialized fields which are unlikely to arise in numerical relativity unless sought specifically.  We do not consider them further here.

The central open issue is this.  At a typical point in the exterior of a black hole formed in a numerical simulation, the Weyl tensor will most likely be of type I.  At late times, perturbation theory is expected to become sufficiently accurate to be meaningfully applied and the perturbed Kinnersley tetrad, when fixed to first order according to Eq.~(\ref{Kinn}), will then belong to a transverse frame of the physical metric.  In order to implement the Teukolsky formalism in this situation, one must discover to which of the three transverse frames of the physical geometry this tetrad belongs.  There are a number of ways one might do this in practice \cite{comp}, but all are based on the same basic idea.  If it is possible to distinguish the eigen-\textit{value} of the Weyl tensor associated with the quasi-Kinnersley frame, then its eigen-\textit{form} can be singled out using Eq.~(\ref{eigen}).  The frame may then be reconstructed from the form using Eq.~(\ref{recon}).

The three eigenvalues of the Weyl tensor at a given space-time point are independent of any basis or coordinate system one might use to evaluate them; they are algebraic scalar curvature invariants of the Weyl tensor.  As such, they should be calculable directly from space-time geometry.  To realize this point explicitly, we calculate the characteristic polynomial 
\begin{equation}\label{poly}
	P := \psi^3 - \tsfrac{1}{4} I\, \psi + \tsfrac{1}{4} J = 0.
\end{equation}
of the linear map defined by the Weyl tensor from the space of self-dual 2-forms to itself.  This calculation is described in some detail in an Appendix.  The roots $\psi$ of this polynomial are the three values of $\Psi_2^\tran$ in the three transverse frames at a generic point of space-time, which are related to the eigenvalues by Eq.~(\ref{eigen}).  The quantities $I$ and $J$ appearing here are the well-known scalar curvature invariants defined by 
\begin{equation}\label{IJdef}\hss
 	\begin{array}{@{}r@{}l@{}}
		I &{}:= \tsfrac{1}{16} \left( C_{ab}{}^{cd} \, C_{cd}{}^{ab} - i \,
			C_{ab}{}^{cd} \,{}^\star\! C_{cd}{}^{ab} \right) \\[1ex]
		J &{}:= \tsfrac{1}{96} \left( C_{ab}{}^{cd} \, C_{cd}{}^{ef} \, C_{ef}{}^{ab} - i \,
			C_{ab}{}^{cd} \, C_{cd}{}^{ef} \,{}^\star\! C_{ef}{}^{ab} \right), 
	\end{array}
\end{equation}
where $\,{}^\star\! C_{ab}{}^{cd} := \frac{1}{2}\, \epsilon_{ab}{}^{mn}\, C_{mn}{}^{cd}$ denotes the space-time dual of the Weyl tensor.  These invariants are manifestly calculable directly from space-time geometry, with no mention of Newman--Penrose tetrads, and show implicitly that the three $\Psi_2^\tran$ are indeed curvature invariants.  This result may be somewhat surprising at first since the curvature component $\Psi_2$ generally depends entirely on the frame used to evaluate it.  However, we seek here its values only in specific, \textit{geometrically privileged}, transverse frames.  Since these frames are defined invariantly, so are the values of $\Psi_2^\tran$ in them.

\begin{figure}
	\includegraphics[width=0.9\columnwidth]{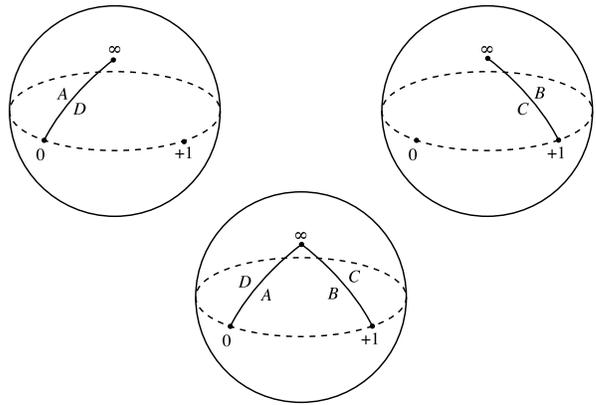}
	\caption{The Riemann surface $\Zsi$ is a triple-cover of the Riemann 
		sphere.  The values of $Z(\si)$ on its three sheets give the 
		three values of $\psi$.  The labels in the figure indicate contiguous 
		regions of $\Zsi$ pictured on different sheets.}\label{surface}
\end{figure}

The cubic polynomial of Eq.~(\ref{poly}) is simple enough to solve analytically, but its solutions will involve multiple-valued functions of the complex variables $I$ and $J$.  To minimize the subtleties raised by these functions, it is convenient to express the solution in terms of the Baker--Campanelli \textsl{speciality index} \cite{BC} $\si := 27 J^2 / I^3$, yielding
\begin{equation}\label{soln}
	\psi = -\frac{3J}{2I}\, Z(\si)
		:= -\frac{3J}{2I}\, \frac{[W(\si)]^{1/3} + [W(\si)]^{-1/3}}{\sqrt{S}},
\end{equation}
where $W(\si) := \sqrt{\si} - \sqrt{\si-1}$.  The multiple values of $\psi$ then arise from the various branches of $Z(\si)$.  This function has branch points of order two at $\si=0$ and $\si=1$ arising from the square roots in $W(\si)$ and, since $W(\si)$ does not vanish for any finite value of $\si$, only one branch point of order three at $\si=\infty$.  A careful analysis shows the Riemann surface $\Zsi$ for $Z(\si)$ has the three-sheeted structure depicted in Fig.~\ref{surface}.  The three eigenvalues of the Weyl tensor arise from the values of $Z(\si)$ in these three sheets.

The speciality index was introduced to provide a  quantitative measure of whether a given physical metric should admit an accurate description in perturbation theory \cite{BC}.  It equals unity if and only if space-time is either of type D or of type II.  (It ill-defined for space-times of types N and III, where $I = 0 = J$, but we do not consider these exotic cases here.  They could be incorporated consistently, however, by working with a multivariate expression for $Z$ in terms of $I$ and $J$.)  One therefore expects $\si \to 1$ at late times in a numerical space-time describing a black hole approaching equilibrium.  Only one branch of $Z(\si)$, associated with the leftmost leaf of $\Zsi$ in Fig.~\ref{surface}, is analytic in this limit.  Let $\psi^0$ denote the value of $\psi$ computed in this branch and expand about $\si = 1$ to find 
\begin{equation}\label{expand0}
	\psi^0 \sim \frac{J}{I} \left[ -3 + \tsfrac{4}{3} (\si-1) + \cdots \right].
\end{equation}
Because $Z(\si)$ is analytic at $\si=1$ in this leaf of $\Zsi$, we find a proper Taylor expansion.  On the other hand, $\si=1$ is a branch point of order two in the other two leaves, and the expansion there should accordingly be done in powers of $\sqrt{\si-1}$ rather than $\si-1$.  Denoting the values of $\psi$ computed in these branches by $\psi^\pm$, we expand to find  
\begin{equation}\label{expandpm}
	\psi^\pm \sim \frac{J}{I} \left[ \tsfrac{3}{2} \pm \tsfrac{i\sqrt{3}}{2} (\si-1)^{1/2} 
		- \tsfrac{2}{3} (\si-1) + \cdots \right]
\end{equation}
These two expansions are completely independent of how the branch lines for $\Zsi$ are drawn in the complex $\si$-plane.  The branch point at $\si=1$ makes it impossible to distinguish the roots $\psi^\pm$ from one another in any neighborhood of that point, and changing the ambiguous sign of $\sqrt{\si-1}$ indeed merely interchanges these two roots.

Since the eigenvalues $\psi^\pm$ are equal when $\si=1$, one expects they must be associated with the degenerate eigen-space of the Weyl tensor on a Kerr space-time, meaning $\psi^0$ must be associated with the Kinnersley frame.  One can check this explicitly by writing the $I$ and $J$ in terms of Newman--Penrose components:
\begin{equation}\label{IJnp}\hss
	\begin{array}{@{}r@{}l@{}}
		I &{}= \Psi_0 \Psi_4 + 3\Psi_2^2 - 4 \Psi_1 \Psi_3 \\[1ex]
		J &{}= \Psi_0 \Psi_2 \Psi_4 - \Psi_2^3 + 2 \Psi_1 \Psi_2 \Psi_3 - 
			\Psi_0 \Psi_3^2 - \Psi_1^2 \Psi_4.
	\end{array}
\end{equation}
In the Kinnersley frame on a type D space-time, only $\Psi_2$ is non-vanishing, and these formulae show that $\Psi_2 = -3J/I$ in that frame.  This, of course, is exactly the value of $\psi^0$ when $\si=1$.  However, the expansion of Eq.~(\ref{expand0}) remains valid within some finite radius of convergence in the complex $\si$-plane.  Within this radius, we can \textit{rigorously} identify a \textit{unique} eigen-form $\Sigma^0_{ab}$ of the Weyl tensor satisfying 
\begin{equation}\label{qKinn}
	C_{ab}{}^{cd}\, \Sigma^0_{cd} = \psi^0\, \Sigma^0_{ab},
\end{equation}
We then \textit{define} the quasi-Kinnersley frame as that associated to $\Sigma^0_{ab}$ by Eq.~(\ref{recon}).  This definition is valid whenever $\si$ is sufficiently close to unity.  

This quasi-Kinnersley frame may be selected in a number of alternative, but equivalent, ways.  Definition 3 of  the companion paper \cite{comp} offers one such alternative, essentially using the distinct limits $\psi^0 \to -3J/I$ and $\psi^\pm \to 3J/2I$ as $\si \to 1$ to distinguish the desired frame.  That procedure will of course select the same quasi-Kinnersley frame described above.  However, we have formulated the definition given here to emphasize a point.  In numerical work, one has at one's disposal only the given physical metric and cannot actually take the limit $\si \to 1$ at a particular point of the numerical space-time.  Nonetheless, for a given $\si \ne 1$, Eq.~(\ref{expand0}) gives an unequivocal value for $\psi^0$ which may be used to select the quasi-Kinnersley frame via Eq.~(\ref{qKinn}).  This branch, and therefore the associated frame, does not exist only in the limit $\si \to 1$, but rather in a finite neighborhood of $\si=1$ in $\Zsi$ given by the radius of convergence of the series of Eq.~(\ref{expand0}).  Thus, although the quasi-Kinnersley frame is selected fundamentally by its limiting behavior, it is defined at points of space-time where $\si$ differs quite substantially from unity.  To understand the limitations of the definition, therefore, we need to analyze the series of Eq.~(\ref{expand0}).  Fortunately, the essential features of the series are evident in the structure of the Riemann surface $\Zsi$.  They are brought out by the following, relatively practical, considerations.

In practice, the computer cannot calculate $\psi^0$ by summing the (infinite) series whose leading terms are given in Eq.~(\ref{expand0}).  This series can be very slow to converge at points.  Rather, it is better to calculate $\psi^0$ non-perturbatively by picking particular branch cuts and branches for the square- and cube-roots in Eq.~(\ref{soln}) such that the series expansion of $\psi$ with those choices yields $\psi^0$ near $\si=1$.  It turns out the principal branches of all three radicals, where $\sqrt{1} = 1 = \sqrt[3]{1}$, with the conventional branch lines along the negative real axis yield $\psi^0$ everywhere in the disk $|\si-1|<1$.  The ambiguous sign of $\sqrt{\si-1}$ causes no problem in this region because changing it, given the branches of the other radicals, merely interchanges the two cube roots in $Z(\si)$.  Thus, $\psi^0$ is continuous across that part of the real axis joining $\si=0$ and $\si=1$.  This is a crude reflection of the absence of any branch line going to $\si=1$ in the corresponding leaf of $\Zsi$.

When $|\si-1|<1$, we have given a simple algorithm to calculate $\psi^0$, and therefore the quasi-Kinnersley frame, non-perturbatively.  Every perturbed Kerr space-time studied by the Teukolsky formalism will certainly satisfy this inequality everywhere outside the black hole.  On such space-times, the quasi-Kinnersley frame provides a gauge-fixed version of the perturbed Kinnersley frame which is manifestly derived solely from the physical metric.  However, other space-times, which in no way represent small perturbations of the Kerr geometry, also admit a quasi-Kinnersley frame.  This structure is therefore uniquely and rigorously defined on every space-time in a \textit{finite} neighborhood of the Kerr solution in the space of solutions to the Einstein equations.

It may be possible to extend the definition of the quasi-Kinnersley frame to even more general regions of space-time using a continuity argument.  The critical ingredient of the definition which fails when $|\si-1|>1$ is that it becomes impossible to distinguish $\psi^0$ from $\psi^\pm$ because of the branch point at $\si=0$ in $\Zsi$.  Concretely, if one continued to use the principal branches of the radicals in Eq.~(\ref{soln}) to define $\psi^0$ in this broader region, this eigenvalue, and therefore the distribution of associated frames, will be discontinuous at any point of a given curve in space-time where $\si$ is real and negative.  This discontinuity, however, stems entirely from the branch line in $\Zsi$.  If we instead define a function $\psi$ along such a curve which equals $\psi^0$ up to the point on the curve where $\si$ is real and negative, and thereafter equals one of the other roots, either $\psi^+$ or $\psi^-$, then $\psi$ \textit{will} allow a continuous extension of the quasi-Kinnersley frame along the curve.  The roots $\psi^\pm$ can be found using the principal branches of the two square roots in Eq.~(\ref{soln}), and the branch of the cube root where $\sqrt[3]{1} = (-1\pm i\sqrt{3})/2$.  In principle, it appears this extension of the quasi-Kinnersley frame to strong-field regions can be carried out coherently (i.e., such that the continuous propagation of the frame from one space-time point to another is independent of the curve used to join them) provided there is no 3-volume in space-time where the curvature is everywhere either type N or III.  In practice, however, if one is interested in harnessing the conceptual power of the Teukolsky formalism in numerical relativity, this may not be worthwhile.  This ``root-substitution'' scheme could incur a high computational expense since it can only be done retroactively, once the late-time evolution is known, and perturbation theory would likely be unreliable anyway when $|\si - 1| > 1$.  Nonetheless, there may be situations where the numerical overhead could be justified.  For example, \cite{BBscal} has advanced a definition of a \textsl{radiation scalar} based on exactly this sort of approach to defining the quasi-Kinnersley frame in strong-field regions of space-time.  That argument yields a scalar invariant $\xi$, defined throughout space-time, which has a clear interpretation in Teukolsky-type radiation zones.

\section{Initial-Value Formulation}

The previous section has defined the quasi-Kinnersley frame from a space-time perspective.  However, this language may not be the most convenient for numerical relativity.  Few numerical codes currently implement a space-time covariant form of the Einstein equations, preferring a space+time split where gravity is encoded in a set of fields evolving relative to a fiducial foliation by space-like hypersurfaces.  These fields include at least the intrinsic spatial metric $h_{ab}$ and extrinsic curvature $K_{ab}$ of each leaf of the foliation.  This section describes a simple way to calculate the quasi-Kinnersley frame using only these initial-value data, thereby offering a direct path toward numerical implementation.

Let $\hat\tau^a$ denote the unit future-directed normal to a given space-like hypersurface in space-time.  Following the convention used in Maxwell theory, a general unit self-dual 2-form $\Sigma_{ab}$ can be decomposed into \textsl{electric} and \textsl{magnetic parts} as\relax
\footnote{Following the convention of Penrose's abstract index notation \cite{Pai}, latin indices are used to denote both space-time and spatial tensors.  The intended meaning should be clear in context.}
\begin{equation}\label{dual}
	\Sigma_{ab} = \hat\sigma_{[a}\, \hat\tau_{b]} + \tsfrac{i}{2}\, \epsilon_{abc}\, \hat\sigma^c.
\end{equation}
Here, $\epsilon_{abc} := \hat\tau^m\, \epsilon_{mabc}$ denotes the intrinsic spatial volume element and $\hat\sigma^a$ is a \textit{complex} spatial vector with real, unit norm.  Since $\Sigma_{ab}$ is self-dual, its electric part (technically, $-\hat\sigma^a/2$) also determines its magnetic part, and therefore $\Sigma_{ab}$ itself, completely.  Thus, at each point on a given spatial hypersurface, there is a one-to-one correspondence between \textit{complex} spatial unit vectors (up to sign) and null frames on space-time at that point.

An equation which determines the $\hat\sigma^a$ corresponding to \textit{transverse} frames follows by taking the electric part of both sides of Eq.~(\ref{eigen}): 
\begin{equation}\label{emred}
	C^a{}_{b}\, \hat\sigma^b := \left( E^a{}_b - i B^a{}_b \right) \hat\sigma^b = 
		2\Psi_2^\tran\, \hat\sigma^a.
\end{equation}
This is once again an eigenvalue problem whose solutions are proportional to the $\Psi_2$ scalar evaluated in a transverse frame.  The spatial tensor $C_{ab}$ featured here is symmetric and trace-free, though generally complex.  Its real and imaginary parts are the electric and magnetic components of the Weyl tensor itself, defined relative to the given spatial slice by 
\begin{equation}\label{emparts}
	\begin{array}{@{}r@{}l@{}}
		E_{ab} &{}:= \hat\tau^m\, \hat\tau^n\, C_{ambn} \\[1ex]
		B_{ab} &{}:= \hat\tau^m\, \hat\tau^n\, {}^\star\! C_{ambn} := 
			\tsfrac{1}{2}\, \hat\tau^m\, \hat\tau^n\, \epsilon_{amij}\, C^{ij}{}_{bn},
	\end{array}
\end{equation}
respectively.  Both of these tensors are obviously real, and the symmetries of the Weyl tensor imply each is separately symmetric and trace-free.

The Gauss--Codazzi--Mainardi relations of differential geometry yield expressions for $E_{ab}$ and $B_{ab}$ in terms of initial-value data.  The result, which applies regardless of whether the space-time metric satisfies the Einstein equations, suggests we consider the quantities 
\begin{equation}\label{emadm}
	\begin{array}{@{}r@{}l@{}}
		\mathcal{E}_{ab} &{}:= {}^3\! R_{ab} + K\, K_{ab} - K_{am}\, K_b{}^m - 
		 	\tsfrac{1}{2}\, h_a^m\, h_b^n\, {}^4\! G_{mn} \\[1ex]
		\mathcal{B}_{ab} &{}:= -\epsilon_a{}^{mn}\, D_m\, K_{nb}.
	\end{array}
\end{equation}
Here, $K := h^{ab}\, K_{ab}$ is the traced extrinsic curvature, while ${}^3\! R_{ab}$ is the Ricci curvature of the torsion-free connection $D_a$ compatible with the spatial metric $h_{ab}$.  The last term in the expression for $\mathcal{E}_{ab}$ is the spatial projection of the \textit{space-time} Einstein tensor.  When the Einstein evolution equations hold, therefore, this term can be replaced with the stress tensor of matter in the spatial slice, and thereby expressed in terms of Cauchy data for the matter fields.  Thus, $\mathcal{E}_{ab}$ and $\mathcal{B}_{ab}$ are functions only of initial-value data when the equations of motion hold.  Moreover, $\mathcal{E}_{ab}$ is naturally symmetric and its trace-free part is $E_{ab}$, while $\mathcal{B}_{ab}$ is naturally trace-free and its symmetric part is $B_{ab}$.  Thus, one can compute the tensor $C_{ab}$ in Eq.~(\ref{emred}) from initial-value data for both the gravitational and matter fields.

The eigenvalue problem of Eq.~(\ref{emred}) is easy enough to solve in practice --- $C_{ab}$ is, after all, only a $3 \times 3$ matrix once coordinates are chosen --- but we must once again pick from among its (generally) three solutions the one corresponding to the quasi-Kinnersley frame.  To do so, we first find  
\begin{equation}\label{IJiv}
	I = \tsfrac{1}{2}\, C^a{}_b\, C^b{}_a \quad\mbox{and}\quad
	J = -\tsfrac{1}{6}\, C^a{}_b\, C^b{}_c \, C^c{}_a.
\end{equation}
These formulas give the fundamental algebraic invariants of the space-time Weyl tensor as functions of spatial initial-value data.  One can then easily compute the speciality index and, provided $|\si-1|<1$,  calculate $\psi^0$ using  Eq.~(\ref{soln}).  As before, $\psi^0$ can be found using the principle branches of the square- and cube-roots in Eq.~(\ref{soln}) with the branch cuts for all three radical functions along the negative real axis.  This preferred eigenvalue determines a preferred eigenvector $\hat\sigma_0^a$ in Eq.~(\ref{emred}) associated to the quasi-Kinnersley frame.

Once the eigenvector $\hat\sigma_0^a$ has been found, one might like to find the elements of some tetrad in the quasi-Kinnersley frame.  This, too, can be done using only spatial, rather than space-time, data.  The real null vectors of an arbitrary tetrad project into a space-like hypersurface to define a pair of real spatial vectors.  While the normalizations of these spatial projections naturally vary if one performs a spin-boost on the tetrad, they define a pair of invariant rays in the tangent space at a point of space-time.  These rays correspond to a pair of real unit vectors $\hat\lambda^a$ and $\hat\nu^a$ which are generally completely independent of one another.  In particular, they point in exactly opposite directions only when the normal $\hat\tau^a$ to the spatial slice lies in the space-time tangent 2-plane spanned by $\ell^a$ and $n^a$.  The fiducial time foliation used in numerical relativity is overwhelmingly likely be ``boosted'' relative to the 2-plane defined by a given null tetrad, and only parallel unit vectors are actually disallowed.

Since $\hat\lambda^a$ and $\hat\nu^a$ determine the directions of the space-time vectors $\ell^a$ and $n^a$, they also suffice to determine the frame associated to $\hat\sigma^a$, and thus $\hat\sigma^a$ itself.  The explicit relation is 
\begin{equation}\label{sln}
	\hat\sigma^a = (1 - \hat\lambda \cdot \hat\nu)^{-1}\, 
		(\hat\lambda^a - \hat\nu^a - i\epsilon^{abc}\, \hat\lambda_b\, \hat\nu_c).
\end{equation}
Note that when $\hat\lambda^a$ and $\hat\nu^a$ are interchanged, $\hat\sigma^a$ simply changes sign, as it should since $\Sigma_{ab}$ changes sign under an exchange operation in the corresponding frame.

To invert Eq.~(\ref{sln}) and solve for $\hat\lambda^a$ and $\hat\nu^a$ given $\hat\sigma^a$, we separate $\hat\sigma^a$ into real and imaginary parts:
\begin{equation}\label{sri}
	\hat\sigma^a = x^a + iy^a.
\end{equation}
In general, neither $x^a$ nor $y^a$ is unit, but the normalization condition for $\hat\sigma^a$ demands they be orthogonal with norms satisfying $\| x \|^2 - \| y \|^2 = 1$.  Taking the electric part of Eq.~(\ref{recon}) then yields
\begin{equation}\label{lns}
	\hat\lambda^a = \frac{x^a + \epsilon^{abc}\, x_b\, y_c}{\| x \|^2} \quad\mbox{and}\quad
	\hat\nu^a = \frac{-x^a + \epsilon^{abc}\, x_b\, y_c}{\| x \|^2}.
\end{equation}
Finally, the elements of the most general null tetrad $(\ell^a, n^a, m^a, \bar m^a)$ in the frame associated to $\hat\sigma^a$ take the form 
\begin{equation}\label{lnproj}
	\begin{array}{@{}r@{}l@{}}
		\ell^a &{}= \frac{|c|}{\sqrt{1 - \hat\lambda \cdot \hat\nu}}\, (\hat\tau^a + \hat\lambda^a) \\[2ex]
		n^a &{}= \frac{|c|^{-1}}{\sqrt{1 - \hat\lambda \cdot \hat\nu}}\, (\hat\tau^a + \hat\nu^a) \\[2ex]
		m^a &{}= \frac{e^{i\theta}}{\sqrt{2}}\, 
			\frac{\sqrt{1 + \hat\lambda \cdot \hat\nu}}{\sqrt{1 - \hat\lambda \cdot \hat\nu}}\,
			(\hat\tau^a + \hat\mu^a),
	\end{array}
\end{equation}
where the \textit{complex} unit projection of $m^a$ is 
\begin{equation}\label{mproj}
	\hat\mu^a = (1 + \hat\lambda \cdot \hat\nu)^{-1}\, (\hat\lambda^a + \hat\nu^a + 
		i \epsilon^{abc}\, \hat\lambda_b\, \hat\nu_c).
\end{equation}
These vectors are the result of a spin-boost, with parameter $c = |c| e^{i\theta}$, acting on a preferred tetrad determined by setting $\ell^a\, \hat\tau_a = n^a\, \hat\tau_a < 0$ and $m^a\, \hat\tau_a = \bar m^a\, \hat\tau_a \le 0$.

Note that in the degenerate case where $\hat\lambda^a$ and $\hat\nu^a$ are indeed anti-parallel, the expression for $m^a$ in Eq.~(\ref{lnproj}) is ill-defined.  However, the limit as $\hat\nu^a \to -\hat\lambda^a$ does exist, though it does depend on how the limit is taken.  Should this situation arise in practice, it is easy enough to accommodate.  The first two of Eqs.~(\ref{lnproj}) are non-singular, and remain unchanged.  One can then choose \textit{any} real unit vector $\hat r^a$ in the spatial 2-plane orthogonal to $\hat\lambda^a = -\hat\nu^a$, and take 
\begin{equation}\label{mexcept}
	m^a = {\textstyle \frac{e^{i\theta}}{\sqrt{2}}} (\hat r^a + i \epsilon^{abc}\, \hat\lambda_b\, \hat r_c).
\end{equation}
This replaces the third of Eqs.~(\ref{lnproj}) to give the remaining two elements of the desired (oriented) null tetrad on space-time.  Note that there are two arbitrary parameters, the unit vector $\hat r^a$ and the spin parameter $\theta$, needed to specify $m^a$ in this degenerate case. They are redundant.  Any rotation of $\hat r^a$ in the plane orthogonal to $\hat\lambda^a$ may be compensated by an adjustment to the spin parameter $\theta$; as in the non-degenerate case, the spin-boost freedom in the tetrad will remain.  The only cost of degeneracy is the loss of the natural ``gauge-fixing'' of the spin freedom implicitly used in Eq.~(\ref{lnproj}).  That is, we cannot specify $\theta = 0$ by insisting that $\hat\tau^a\, m_a$ be real and negative since, in the degenerate case, it vanishes for all values of $\theta$.  In any event, this amounts only to a slight mathematical inconvenience.  The residual freedom in the tetrad is the same regardless of whether $\hat\lambda^a = -\hat\nu^a$ and, moreover, this circumstance is unlikely to be realized \textit{exactly} in numerical relativity.

This section has now shown explicitly that transverse frames can be calculated directly from data on an \textit{arbitrary} spatial slice of a given space-time.  In particular, one can find these space-time quantities using the fiducial foliation underlying any numerical evolution scheme based on the familiar Arnowit--Deser--Misner equations or their various descendants.

\section{Conclusions}

The central point this paper has argued is that the quasi-Kinnersley frame is naturally defined on a broad class of space-times.  As long as space-time contains no finite volumes of very high algebraic speciality, a ``root-substitution'' scheme can propagate the quasi-Kinnersley frame continuously from regions supporting fairly weak fields, where there is absolutely no ambiguity in its definition, to arbitrary regions supporting much stronger fields.  To avoid potential subtleties inherent to this scheme, one can restrict attention to situations where $|\si - 1| < 1$ throughout space.  In this restricted case, the quasi-Kinnersley frame is particularly easy to find.  In the still more restrictive limit $\si \to 1$, the quasi-Kinnersley frame always exists, depends only on the physical metric, and identifies a ``gauge-fixed'' version of the perturbed Kinnersley frame satisfying Eq.~(\ref{Kinn}).  This provides essential guidance needed to deploy the Teukolsky formalism where it is expected to be relevant.  This feature could be useful in extracting geometrically-invariant, \textit{physically meaningful} information from the raw numerical data of late-time black hole simulations.

The mathematical structure discussed above could also be useful in constructing observer-independent scalar functions in more general regions where perturbation theory  may not apply.  Although the physical meaning of these scalars may be somewhat obscure in regions supporting self-interacting fields, they can still be used, in principle, as gauge-invariant observables and thus help clarify invariantly the structure of simulated space-times.

It does not seem possible to establish concrete, analytic limits on the regime where perturbative techniques are justified.  Their viability must be tested by some other means.  The internal consistency checks of the Lazarus project \cite{Laz}, where late-time evolution is done by extracting Teukolsky quantities from numerical data on a given time-slice in a non-linear evolution and thereafter using the Teukolsky formalism's much more stable evolution equations, provide a natural way to do this.  These consistency checks ensure the physical content of the very late-time data does not depend sensitively on when the extraction took place.  This paper has argued that one can begin attempting extractions whenever one first finds $|\si - 1| < 1$ everywhere outside a single black hole.  In practice, it appears this can happen surprisingly early in a full numerical evolution \cite{Laz}.

In addition to transverse frames, it is possible to consider several other geometrically-preferred frames on a typical space-time.  For example, frames whose real null directions are two of the (generally four) principal null directions of the physical metric are quite natural to consider.  The connection between such principle null frames and the transverse frames considered here is discussed much more thoroughly in the companion paper \cite{comp} (see especially section II.F).  One might wonder whether techniques similar to those outlined here for the quasi-Kinnersley frame could be developed to pick some of these other preferred frames on a generic geometry.  The answer, however, is probably in the negative.  The essential feature which has distinguished the quasi-Kinnersley frame in this paper is the analyticity of $\psi^0$ near $\si = 1$.  Even the other two transverse frames cannot be distinguished from one another near algebraic speciality because of the branch cut joining their associated leaves of the Riemann surface at this point.  One can show using standard perturbation theory of linear operators that, since $\psi^\pm$ fail to vary analytically near speciality, their associated frames also fail to do so \cite{Mup}.  This non-analyticity, and our inability to distinguish these frames invariantly, arises precisely because we have perturbed away from a situation where the normally distinct frames degenerate.  All of the other obvious geometrically-preferred frames, such as those based on principal null directions, also degenerate in algebraically special space-times, and thus will share this problem.  The quasi-Kinnersley frame is \textit{unique} in that it varies analytically when perturbed, and is therefore arguably a singularly useful structure in the sort of analyses contemplated here.

As discussed in the Introduction, a second step must be taken to complete the construction begun in this paper and allow an application of the full range of Teukolsky's techniques to the results of numerical simulations.  This step will break the spin-boost-invariance of the current results and choose a preferred \textsl{quasi-Kinnersley tetrad} within the quasi-Kinnersley frame.  Unlike the frame-fixing procedure described above, which has been entirely local, done separately at each space-time point, fixing the scalings seems likely to require a \textit{global} approach.  Setting $\epsilon = 0$, as for the Kinnersley tetrad on Kerr space-time, can still be done generically, and will continue to help pare down the possibilities for a quasi-Kinnersley \textit{tetrad}.  However, as in Kinnersley's case, this will not determine the tetrad uniquely.  In fact, since the $\epsilon = 0$ condition ties together the scalings of tetrad elements along the integral curves of $\ell^a$, it would seem that fixing the scaling close to the black hole would involve making choices in the evolved fields at the outer boundary at \textit{much} later times.  Since long-time evolutions of the full, non-linear equations of general relativity are currently unavailable, it seems likely an indirect approach to this problem will be required for the near-term.  It may be possible to found such a approach on a scheme to approximate additional structures of the Kerr geometry underlying a given perturbed metric using only the physical metric itself.  Work is currently underway to explore whether such an approach could yield a practicable solution to the problem of finding the quasi-Kinnersley tetrad.

\begin{acknowledgments}

The authors would like to thank Richard Price for several stimulating discussions of this work.  CB would also like to thank Manuela Campanelli, Karel Kucha\v r, Alok Laddha, Carlos Lousto and Warner Miller for helpful conversations, and the NASA Center for Gravitational Wave Astronomy at University of Texas at Brownsville (NAG5-13396) for its hospitality during part of the execution of this work.  This work was supported by the National Science Foundation through grants PHY-0244605 to the University of Utah and PHY-0400588 to Florida Atlantic University, as well as by the EU Network Programme through Research Training Network contract HPRN-CT-2000-00137 and NASA through grant NNG04GL37G to the University of Texas at Austin.

\end{acknowledgments}

\appendix

\section{The Characteristic Polynomial}

The characteristic polynomial of Eq.~(\ref{poly}) plays a key role in identifying the quasi-Kinnersley frame.  For definiteness, this Appendix will derive it.  The derivation follows closely the discussion given in Chapter 4 of \cite{exact}.

The space of self-dual 2-forms at a point of space-time is a three-dimensional complex vector space.  To see this, we recall that in Lorentzian signature, any 2-form $\Omega_{ab}$ may be written as a sum of its self-dual and anti-self-dual parts:
\begin{equation}\label{decomp}
	\Omega_{ab} = {}^+ \Omega_{ab} + {}^- \Omega_{ab}, 
\end{equation}
where ${}^\pm \Omega_{ab} := \frac{1}{2} \left[ \Omega_{ab} \mp i\,{}^\star\Omega_{ab} \right]$ are the \textsl{self-dual} (upper sign) and \textsl{anti-self-dual} (lower sign) \textsl{parts} of $\Omega_{ab}$.  One can check the 2-form ${}^+\!\Omega_{ab}$ (${}^-\!\Omega_{ab}$) is (anti-)self-dual using the result ${}^{\star\star}\Omega_{ab} := {}^\star({}^\star\Omega_{ab}) = -\Omega_{ab}$, which holds for the double dual of any 2-form in Lorentzian signature.  The space of all 2-forms in four dimensions is six dimensional, but the decomposition of Eq.~(\ref{decomp}) splits it into two subspaces which are complex conjugates of one another.  That is, we see by taking the conjugate of the definition of self-duality in Eq.~(\ref{sd}) that the complex conjugate of a self-dual 2-form is anti-self-dual.  These two subspaces cannot intersect and together form the whole of the space of 2-forms, whence each must be three-dimensional.

The contraction of any self-dual 2-form $\Sigma_{ab}$ with the Weyl tensor will yield another self-dual 2-form.  This happens because the Weyl tensor has the remarkable property that its duals on either pair of indices are equal to one another:
\begin{equation}\label{dualeq}
	{}^\star\!C_{ab}{}^{cd} = C^\star{}_{ab}{}^{cd} := 
		{\textstyle\frac{1}{2}}\, C_{ab}{}^{mn}\, \epsilon_{mn}{}^{cd}.
\end{equation}
Thus, the dual of the contraction $C_{ab}{}^{cd}\, \Sigma_{cd}$ can be moved to act on the second pair of Weyl indices and then to act on $\Sigma_{cd}$, producing a factor of $+i$ and showing the contraction to be self-dual.  Thus, the Weyl tensor defines a linear map from a three-dimensional complex vector space to itself.

In Eq.~(\ref{dual}), we have shown explicitly how the three-dimensional space of self-dual 2-forms $\Sigma_{ab}$ at a point of space-time can be described in terms of complex spatial vectors $\sigma_a := \Sigma_{ab}\, \hat\tau^b$ in an arbitrary spatial hypersurface.  These later objects also form a three-dimensional complex vector space.  Indeed, the map induced by the Weyl tensor on self-dual 2-forms is completely equivalent to the map $C^a{}_b$ on complex spatial vectors defined by Eq.~(\ref{emred}).  For conceptual simplicity, then, we may calculate the characteristic polynomial of the spatial map $C^a{}_b$ rather of than the space-time Weyl tensor restricted to self-dual 2-forms.

For the moment, let $C^a{}_b$ be \textit{any} map on a three-dimensional vector space.  Using standard techniques in linear algebra, one can show the its characteristic polynomial can be expressed in terms of the traces of $C^a{}_b$ and its powers in the form 
\begin{equation}\label{genpoly}
	\begin{array}{r@{}l}
		0 = P(c) &{}= \det \left[ C - c\delta \right] \\[1ex]
			&{}= c^3 - \langle C \rangle\, c^2 + 
				{\textstyle\frac{1}{2}} \left[ \langle C \rangle^2 - \langle C^2 \rangle \right] c \\[1ex]
			&\qquad{} - {\textstyle\frac{1}{6}} \left[ \langle C \rangle^3 - 
				3 \langle C \rangle \langle C^2 \rangle + 2 \langle C^3 \rangle \right],
	\end{array}
\end{equation}
where $\delta^a{}_b$ is the usual identity map and $\langle C^n \rangle$ denotes the trace of the $n^{\mathrm{th}}$ power of $C^a{}_b$.  Returning now to the particular case of the Weyl tensor, we see immediately that these coefficients simplify considerably because $C^a{}_b$ is trace-free.  Moreover, Eqs.~(\ref{IJiv}) relate the remaining traces to the curvature invariants $I$ and $J$.  This reduces the characteristic equation for the eigenvalues of the Weyl tensor to the form 
\begin{equation}\label{weylpoly}
	c^3 - I\, c + 2J = 0.
\end{equation}
Since, by Eq.~(\ref{emred}), the eigenvalue $c$ is twice $\Psi_2^\tran$ (in a transverse frame), we recover the polynomial of Eq.~(\ref{poly}) for the values of $\psi$.

 \end{document}